\documentclass[conference]{IEEEtran}
\IEEEoverridecommandlockouts
\usepackage{cite}
\usepackage{amsmath,amssymb,amsfonts}
\usepackage{algorithmic}
\usepackage{graphicx}
\usepackage{textcomp}
\usepackage{xcolor}
\usepackage{upgreek}
\usepackage{siunitx}
\usepackage{booktabs}
\usepackage{array}
\usepackage{multirow}
\usepackage{microtype} % 自动微调字符间距，消除文字溢出边缘（Overfull \hbox）并让段落边缘更平整
\usepackage{flushend}  % 自动切齐最后一页的左右两栏，避免出现左长右短的情况

\def\BibTeX{{\rm B\kern-.05em{\sc i\kern-.025em b}\kern-.08em
    T\kern-.1667em\lower.7ex\hbox{E}\kern-.125emX}}

\renewcommand{\arraystretch}{1.1} % 将表格行间距稍微拉开 10%，让数据不再拥挤，提升阅读体验

\begin{document}

\title{Decoupling Error Attribution in Cloud-Native Graph-RAG: A Data Integrity Diagnostic Framework\\
\thanks{This research was funded by Science and Technology Projects of Xizang Autonomous Region, China, project title "An Intelligent Question-Answering System for the Ecological Environment of the Qinghai-Tibet Plateau Based on Large Language Model-Powered Knowledge Graph Agents", grant number XZ202502ZY0073.}
}

\author{%
\begin{tabular}{c@{\hspace{2.0em}}c}
\begin{tabular}{@{}c@{}}
1\textsuperscript{st} Shuai Yan*\\
\textit{College of Computer and Software}\\
\textit{Chengdu Jincheng College}\\
Chengdu 611731, China\\
yanshuai1@cdjcc.edu.cn\\
*Corresponding author
\end{tabular}
&
\begin{tabular}{@{}c@{}}
2\textsuperscript{nd} Yuhang Wu\\
\textit{College of Computer and Software}\\
\textit{Chengdu Jincheng College}\\
Chengdu 611731, China\\
wuyuhang1@cdjcc.edu.cn
\end{tabular}
\\[4.4em]
\multicolumn{2}{c}{%
\hspace*{-3.5em}%
\begin{tabular}{c@{\hspace{2.0em}}c}
\begin{tabular}{@{}c@{}}
3\textsuperscript{rd} Xiaodong Huang\\
\textit{College of Computer and Software}\\
\textit{Chengdu Jincheng College}\\
Chengdu 611731, China\\
huangxiaodong@cdjcc.edu.cn
\end{tabular}
&
\begin{tabular}{@{}c@{}}
4\textsuperscript{th} Ke Wang\\
\textit{College of Computer and Software}\\
\textit{Chengdu Jincheng College}\\
Chengdu 611731, China\\
wangke@cdjcc.edu.cn
\end{tabular}
\end{tabular}%
}
\end{tabular}
}

\maketitle

\begin{abstract}
Graph-RAG systems often assume pristine data quality, overlooking the severe impact of perturbations in cloud-native databases. This paper proposes a three-layer decoupled diagnostic framework to orthogonally attribute system errors to reasoning loss, Knowledge Graph (KG) defects, and Cypher generation errors. Evaluated on a spatio-temporal ecological KG of the Southeastern Tibet region with eight defect types, results reveal that data integrity—rather than algorithmic reasoning—is the dominant performance bottleneck, with structural defects degrading system accuracy from 0.93 to 0.39. Crucially, we observe a masking-like phenomenon termed the Parametric Knowledge Masking Effect (PKME), suggesting LLMs compensate for broken retrieval paths using internal memory. This shrinks apparent query generation errors by over 70\%, obscuring actual storage deterioration and increasing the risk of false negatives for automated monitoring. This work provides a quantitative foundation for auditing and optimizing data integrity in cloud-based information fusion systems.
\end{abstract}

\begin{IEEEkeywords}
Graph-RAG; Data Integrity; Error Attribution; Cloud-Native Graph Databases; Parametric Knowledge Masking Effect; Quality Assurance
\end{IEEEkeywords}

\section{Introduction}
In the era of cloud computing and big data, the integration of multi-modal datasets---such as spatio-temporal ecological remote sensing data---into Knowledge Graphs (KGs) has become a fundamental infrastructure for advanced information fusion and retrieval systems~\cite{b7}. However, real-world data integration inevitably introduces critical challenges related to data integrity and quality control. In cloud-native graph databases, data ingestion pipelines across distributed storage nodes inevitably introduce critical challenges such as multi-replica synchronization delays (causing temporal drift) and distributed write conflicts (leading to structural noise)~\cite{b5}. High-quality graph data assets are the critical foundation for these advanced systems to function effectively.

Recently, Knowledge Graph-augmented Retrieval-Generation (Graph-RAG) has emerged as a novel paradigm for data retrieval and integration, utilizing Large Language Models (LLMs) to process structured graph data~\cite{b1,b2,b3,b4,b8,b11}. However, a critical gap persists in contemporary Graph-RAG evaluation: when the system fails or performance degrades, traditional End-to-End (E2E) testing cannot distinguish whether the failure originates from the query engine (e.g., Cypher generation errors~\cite{b12}) or the collapse of the underlying KG data assets. Most existing studies assume that the knowledge graph is of pristine quality~\cite{b1,b2}, overlooking the severe impact of data perturbations. This entanglement of error sources makes it nearly impossible for data engineers to accurately audit data quality and optimize cloud-based graph storage. 

In cloud storage environments and distributed graph databases, data quality perturbations are the norm rather than the exception. To address this critical gap in data understanding and quality assurance from a data management perspective, this paper proposes a systematic data quality diagnostic framework. To the best of our knowledge, this work represents an early attempt to systematically decouple error sources in Graph-RAG strictly from a data integrity perspective. The main contributions are as follows:

(1) We propose a decoupled data quality audit framework that orthogonally decomposes system errors, isolating quantitative losses caused purely by underlying data defects ($\varepsilon_{\text{KG}}$).

(2) We construct a dynamic ecological knowledge graph for the Southeastern Tibet region of the Qinghai-Tibet Plateau, grounded in real-world data to provide a robust benchmark for controlled defect-injection analysis.

(3) We observe a non-monotonic error propagation behavior, suggesting a Parametric Knowledge Masking Effect (PKME) that poses a significant observability challenge by obscuring true data degradation.

\section{Related Work}
Existing research on Knowledge Graph quality management primarily focuses on static rule-based anomaly detection, schema validation, and logical consistency checking~\cite{b5}. While providing foundational integrity constraints, these conventional methods lack the capability to dynamically assess how localized storage defects propagate through complex retrieval pipelines in cloud environments.

In Graph-RAG systems~\cite{b2,b3,b4,b8,b11}, current evaluation paradigms remain predominantly End-to-End (E2E), focusing heavily on final generative outputs or hallucination rates~\cite{b6,b9,b10}. Mainstream LLM evaluation frameworks (such as RAGAS or ARES) primarily measure generation faithfulness and answer relevance at the unstructured text level. Because these black-box pipelines lack structural graph awareness, they fail to perform fine-grained error attribution, leaving Database Administrators (DBAs) unable to distinguish algorithmic query limitations from underlying database corruption. Furthermore, existing paradigms generally treat the KG as a static, pristine asset, often overlooking the degraded data conditions prevalent in real-world cloud deployments.

\section{Dataset and System Configuration}
\subsection{Spatio-Temporal Knowledge Graph Construction}
We construct a dynamic spatio-temporal knowledge graph utilizing ecological remote sensing data from the Southeastern Tibet region of the Qinghai-Tibet Plateau. To achieve robust data interoperability, the graph fuses multi-source heterogeneous data (NDVI, AGB, DEM) through spatial grid aggregation and cross-year temporal linking (2015 and 2021), resulting in a dense topology. The final dynamic knowledge graph contains \num{1,172} nodes (390 GridCells, 2 TimePoints, 780 Observations) and \num{3,428} relationships (780 HAS\_OBSERVATION, 780 IN\_YEAR, 1,478 ADJACENT\_TO, 390 TEMPORAL\_NEXT). This intentionally condensed, high-density scale facilitates exhaustive, fine-grained logical integrity auditing, providing a rigorously controlled foundation for subsequent evaluations.

\subsection{Question-Answering Dataset Design}
An evaluation dataset of 100 question-answering items is constructed through manual design and programmatic verification, covering four representative retrieval scenarios: single-grid attribute queries (30 items), cross-year dynamic comparison (25 items), spatial neighborhood multi-hop reasoning (25 items), and complex spatio-temporal joint reasoning (20 items). The dataset spans Easy to Hard difficulty levels, with 54\% of questions requiring 3-hop graph retrieval.

For each question, a standard Cypher query (gold\_cypher) is manually written and verified on the original clean graph to ensure baseline data integrity. The gold answer (gold\_answer) and gold subgraph (gold\_subgraph) are also recorded. The evaluation adopts an LLM-as-Judge protocol, using Qianwen-Max (temperature = 0) as the judge model. Answers are scored according to nine rule categories (e.g., factual correctness, evidence completeness, and logical consistency) using binary labels (0 or 1), while ranking questions are allowed to receive a partial score of 0.5. This design improves the objectivity and granularity of the evaluation in complex data environments.

\subsection{Controlled Defect Injection Scheme}
To rigorously evaluate the system's resilience under degraded data conditions, we implement a systematic Data Quality Stress Testing mechanism instead of simple noise addition. Using the original clean graph as the baseline, eight independent defective graphs are generated through controlled cloning to simulate real-world database failures. Each graph contains one specific defect type injected at a fixed rate of 20\%. It is important to note that this high injection rate is intentionally selected as a Stress Testing methodology to explore the system's boundary conditions and error propagation limits, rather than modeling daily operational averages. In distributed graph storage, localized infrastructure failures—such as partial shard loss or transient synchronization lags—often manifest as high-density defect bursts. This \num{20}\% injection rate approximates these extreme cloud-native scenarios, providing a necessary stress test for data integration robustness.

Crucially, to simulate enterprise-level database management systems (DBMS) practices common in cloud deployments, database-level logical and physical separation is implemented through Neo4j's multi-database mechanism. While this ensures strict data isolation across experimental conditions, it focuses on data integrity rather than performance isolation. As shown in Table~\ref{tab:defect_types}, the eight defect types mapped here represent logical manifestations of physical cloud storage issues. Structural defects directly alter the graph topology (e.g., deleting or redirecting relations), simulating broken connectivity caused by severe infrastructure events like distributed shard loss or synchronization delays. Conversely, numeric defects corrupt node attributes (e.g., value substitution), reflecting data ingestion anomalies in ETL pipelines. This taxonomy follows mainstream dimensions in knowledge graph quality assessment~\cite{b5}, ensuring both representativeness and methodological rigor for database evaluation.

\begin{table}[t]
\centering
\caption{Eight Controlled Defect Types}
\label{tab:defect_types}
\small
\setlength{\tabcolsep}{3pt}
\renewcommand{\arraystretch}{1.4}
\begin{tabular}{>{\raggedright\arraybackslash}m{2.9cm}>{\centering\arraybackslash}m{1.5cm}>{\raggedright\arraybackslash}m{3.8cm}}
\toprule
\textbf{Defect Type} & \textbf{Category} & \textbf{Primary Impact \& Cloud Mapping} \\
\midrule
Structural Sparsity   & Structural & Query path interruption (e.g., shard loss) \\
Structural Noise      & Structural & Retrieval results point to incorrect entities \\
Temporal Drift        & Structural & Temporal logic confusion (e.g., sync delays) \\
Evidence Gap          & Structural & Evidence exists but content is missing \\
\midrule
Motif Artifact        & Numeric    & Uniform value distribution, misleading statistical reasoning \\
Factual Noise         & Numeric    & Attribute values exceed ecologically plausible ranges \\
Logical Contradiction & Numeric    & Self-contradictory attributes within an entity \\
Numeric Corruption    & Numeric    & Numerical precision and constraint violation \\
\bottomrule
\end{tabular}
\end{table}

\section{Three-Layer Decoupled Evaluation Framework}
\subsection{Design Motivation}
\begin{figure*}[t]
    \centering
    \includegraphics[width=\textwidth]{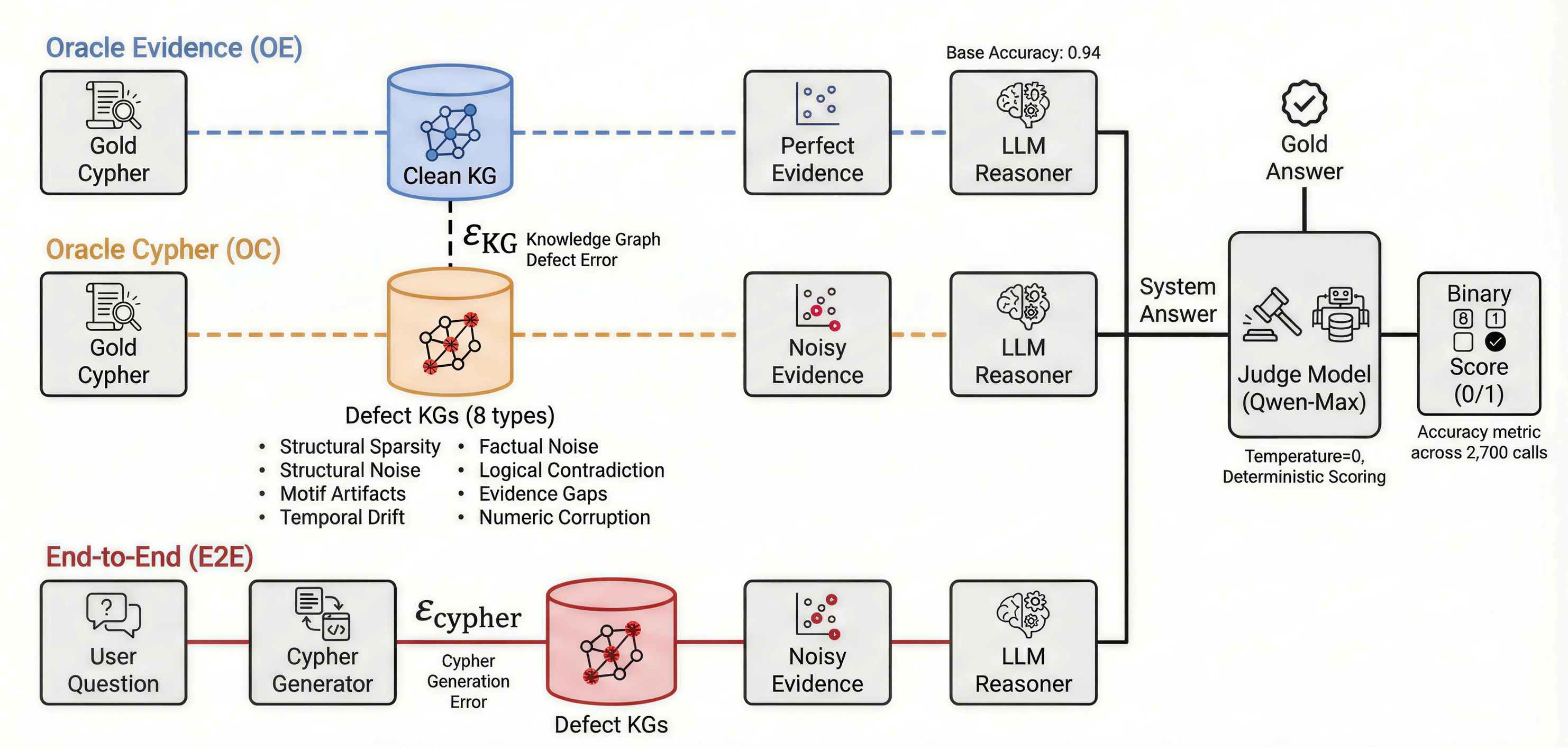}
    \caption{System Architecture and Three-Layer Decoupled Evaluation Flowchart}
    \label{fig1}
\end{figure*}

A key limitation of End-to-End (E2E) evaluation in modern cloud-based data systems is that, when system performance deteriorates, it is difficult to determine whether the loss is caused by the underlying knowledge graph data quality, the query engine's schema alignment, or the reasoning ability of the large language model. To overcome this black-box limitation and provide quantifiable metrics for database administrators, this paper adopts a control-variable strategy to decouple these three error sources. The basic idea is to keep part of the data retrieval pipeline fixed while activating only one source of error at a time. In this way, each evaluation layer isolates a distinct error component, elevating the evaluation process from a simple benchmarking procedure into a structured, white-box error attribution framework for data quality diagnosis. As illustrated in Fig.~\ref{fig1}, the three-layer framework introduces control at different stages of the data integration pipeline and forms an orthogonal evaluation scheme.

\subsection{Three-Layer Framework Definition}
The proposed framework consists of three diagnostic layers, as summarized in Table~\ref{tab:three_layer_def}. The Oracle Evidence (OE) layer uses the standard Cypher query (gold\_cypher) on the clean graph and converts the retrieved result into perfect textual evidence for the LLM. This layer measures the upper bound of downstream reasoning performance under ideal data integrity conditions. The Oracle Cypher (OC) layer executes gold\_cypher on a defective graph to quantify the pure impact of KG defects on retrieval path integrity. The End-to-End (E2E) layer requires the LLM to generate the Cypher query and perform retrieval on a defective graph. This layer audits the performance of the complete information fusion system under realistic, degraded data scenarios. 

\begin{table}[t]
\centering
\caption{Definition of the Three-Layer Decoupled Evaluation Framework}
\label{tab:three_layer_def}
\small
\setlength{\tabcolsep}{3pt}
\renewcommand{\arraystretch}{1.4}
\begin{tabular}{>{\raggedright\arraybackslash}m{2.3cm}>{\raggedright\arraybackslash}m{3cm}>{\raggedright\arraybackslash}m{3cm}}
\toprule
\textbf{Layer} & \textbf{Experimental Design} & \textbf{Measurement Goal} \\
\midrule
Oracle Evidence & gold\_cypher $\rightarrow$ Clean Graph $\rightarrow$ Perfect Text Evidence $\rightarrow$ LLM & Upper bound of LLM reasoning capability under ideal data provenance \\
Oracle Cypher & gold\_cypher $\rightarrow$ Defective Graph $\rightarrow$ Graph Retrieval Results $\rightarrow$ LLM & Impact of KG quality defects on storage integrity and retrieval paths \\
End-to-End & Generated Cypher $\rightarrow$ Defective Graph $\rightarrow$ Graph Retrieval Results $\rightarrow$ LLM & Performance of the complete system under real data defect scenarios \\
\bottomrule
\end{tabular}
\end{table}

\subsection{Error Attribution Formulas}
Let $\text{OC}_{\text{clean}}$ denote the baseline accuracy obtained by using $\text{gold\_cypher}$ on the clean graph, which is measured as 0.93 in this study. For any defective graph, the total system error can then be decomposed into three orthogonal diagnostic components. For a given evaluation layer $\lambda \in \{\text{OE},\text{OC},\text{E2E}\}$ and database $D$, the layer accuracy $\text{Acc}(\lambda, D)$ is defined as the mean binary score $\frac{1}{|Q|} \sum_{q \in Q} s_{q}^{\lambda,D}$ over the question set $Q$, where $s_{q}^{\lambda,D} \in \{0, 0.5, 1\}$ is the judge-assigned score for question $q$ under layer $\lambda$ on database $D$. The decoupled error attribution formulas defined below constitute the core of this data audit mathematical model, where OE and $\text{OC}_{\text{clean}}$ are evaluated on the clean graph and OC, E2E are evaluated on defective graphs.

Based on the evaluations, the total error $\varepsilon_{\text{total}} = \text{OE} - \text{E2E}$ can be orthogonally decomposed into three diagnostic components to provide a clear mapping for database administrators:
1) Reasoning Loss ($\varepsilon_{\text{reason}} = \text{OE} - \text{OC}_{\text{clean}}$) reflects the algorithmic bottleneck; 2) KG Defect Error ($\varepsilon_{\text{KG}} = \text{OC}_{\text{clean}} - \text{OC}$) captures storage integrity loss; and 3) Cypher Gen. Error ($\varepsilon_{\text{cypher}} = \text{OC} - \text{E2E}$) reflects schema alignment failure.

By substituting these isolated components, we derive the orthogonal decomposition identity:
\begin{equation}
\varepsilon_{\text{total}} = \varepsilon_{\text{reason}} + \varepsilon_{\text{KG}} + \varepsilon_{\text{cypher}}
\label{eq:eps_decomp}
\end{equation}
A fundamental assumption of this decomposition is that the LLM's intrinsic reasoning capability remains independent of retrieved context quality. Since the OE layer is evaluated on a pristine graph with gold evidence, it establishes a practical upper bound for the LLM's reasoning capability on a specific query. By fixing this ceiling, the baseline $\varepsilon_{\text{reason}}$ remains an invariant reference point, ensuring that subsequent shifts in $\varepsilon_{\text{KG}}$ and $\varepsilon_{\text{cypher}}$ strictly reflect the damage caused by data integrity degradation. By construction, $\varepsilon_{\text{reason}} \geq 0$ and $\varepsilon_{\text{KG}} \geq 0$. Empirically, $\varepsilon_{\text{cypher}} \geq 0$ always holds, since E2E never outperforms OC on defective graphs.

However, $\varepsilon_{\text{cypher}}$ is not constant: it varies substantially with defect severity, revealing the Parametric Knowledge Masking Effect (PKME). To quantitatively measure this observability risk, we define the Masking Index $MI(D)$ as:
\begin{equation}
MI(D) = \frac{\varepsilon_{\text{cypher}}(\text{clean}) - \varepsilon_{\text{cypher}}(D)}{\varepsilon_{\text{cypher}}(\text{clean})} \times 100\%
\label{eq:mi}
\end{equation}

Empirically, $\varepsilon_{\text{cypher}}$ reaches 0.18 on the clean graph, but shrinks to 0.01--0.05 under severe structural defects. For instance, under Evidence Gap, $MI$ reaches $72.2\%$. We hypothesize that when Cypher queries fail because database retrieval paths are broken, the LLM falls back to its internal parametric knowledge to generate an answer. Importantly, this compensation does not eliminate the underlying data defect ($\varepsilon_{\text{KG}}$); it merely masks the apparent contribution of schema misalignment, creating a critical blind spot for automated data quality monitoring scripts.

\subsection{Statistical Analysis Methods}
The significance threshold is set to $\alpha = 0.05$. Effect size is measured by Cohen's $d$ (computed as the standardized mean difference $d = \mu_{\Delta}/\sigma_{\Delta}$ of the paired score vector), where $d \geq 0.8$ indicates a large effect. A positive $d$ indicates that the first evaluated layer outperforms the second on average.

The experimental system configuration is as follows. Qianwen-Flash (temperature = 0.2) is used for Cypher generation, Qianwen-Plus is used for question answering with extended reasoning enabled, and Qianwen-Max (temperature = 0) is used as the judge model. The data audit system runs with 40 concurrent worker threads. In total, 2,700 inference calls are executed, calculated as 9 databases $\times$ 100 queries $\times$ 3 layers.

\section{Experimental Results}
\subsection{Performance Boundary Analysis}
\begin{figure}[htbp]
    \centering
    \includegraphics[width=0.98\columnwidth]{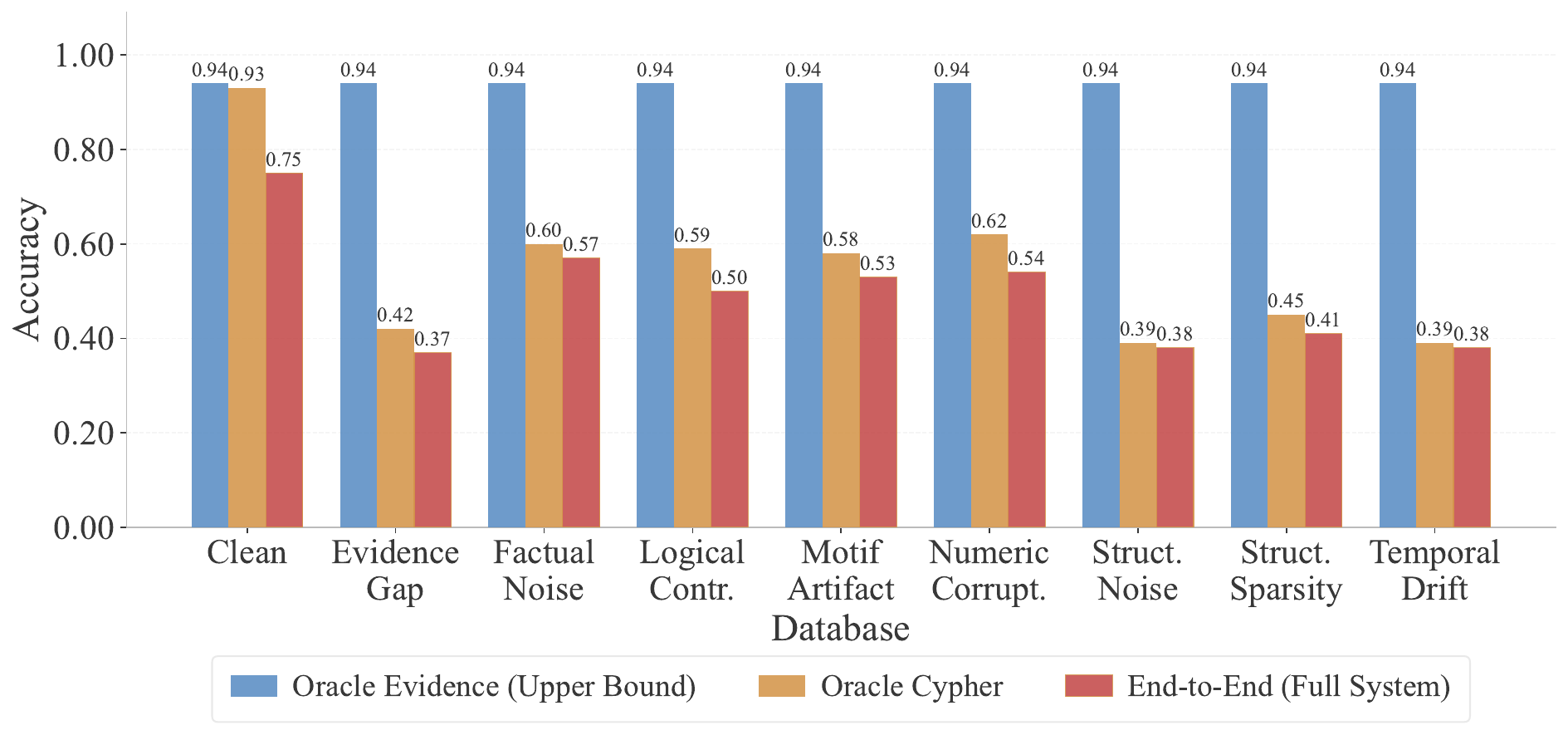}
    \vspace{0.3cm}
    \caption{Three-Layer Baseline Accuracy Comparison}
    \label{fig2}
\end{figure}

Fig.~\ref{fig2} compares the accuracy of the three-layer data audit framework across nine database settings. In the baseline condition, Oracle Cypher on the clean graph (OC\_clean) achieves an accuracy of 0.93, while Oracle Evidence (OE) reaches 0.94. The difference is only 0.01. According to the Wilcoxon signed-rank test, this gap is not statistically significant ($p = 0.706$). This suggests an important implication for data management in similar structured reasoning tasks: the data quality bottleneck may outweigh the algorithmic bottleneck. High-quality structured data assets, rather than the intrinsic reasoning capabilities of the models, predominantly determine the upper bound of system performance. The structured graph retrieval results provide support equivalent to perfect textual evidence for downstream LLM reasoning.

Further analysis reveals substantial heterogeneity across defect types. Structural defects, including Structural Sparsity, Structural Noise, Temporal Drift, and Evidence Gap, reduce Oracle Cypher accuracy to the range of 0.39--0.45, corresponding to declines of more than 50\% relative to OC\_clean. In contrast, numeric defects, including Motif Artifact, Factual Noise, Logical Contradiction, and Numeric Corruption, keep OC accuracy in the range of 0.58--0.62, with declines limited to 33\%--38\%. Mechanistically, structural defects directly break the topological connectivity of database retrieval paths, whereas numeric defects mainly affect the reliability of node attributes without destroying graph reachability. This provides a clear directive for database administrators (DBAs): prioritizing the preservation of topological connectivity during data ingestion and cloud storage optimization is paramount.

Compared with the clean-graph condition, where $\varepsilon_{\text{cypher}} = 0.18$, two severe structural defect scenarios show much smaller Cypher generation error: Evidence Gap ($\text{OC} = 0.42$, $\text{E2E} = 0.37$, $\varepsilon_{\text{cypher}} = 0.05$) and Structural Sparsity ($\text{OC} = 0.45$, $\text{E2E} = 0.41$, $\varepsilon_{\text{cypher}} = 0.04$). These results provide initial evidence for the Parametric Knowledge Masking Effect, whose critical impact on data provenance is examined further in Section~\ref{sec:ivb}.

\subsection{Bottleneck Diagnosis and Error Attribution}
\label{sec:ivb}
\begin{figure}[htbp]
    \centering
    \includegraphics[width=0.98\columnwidth]{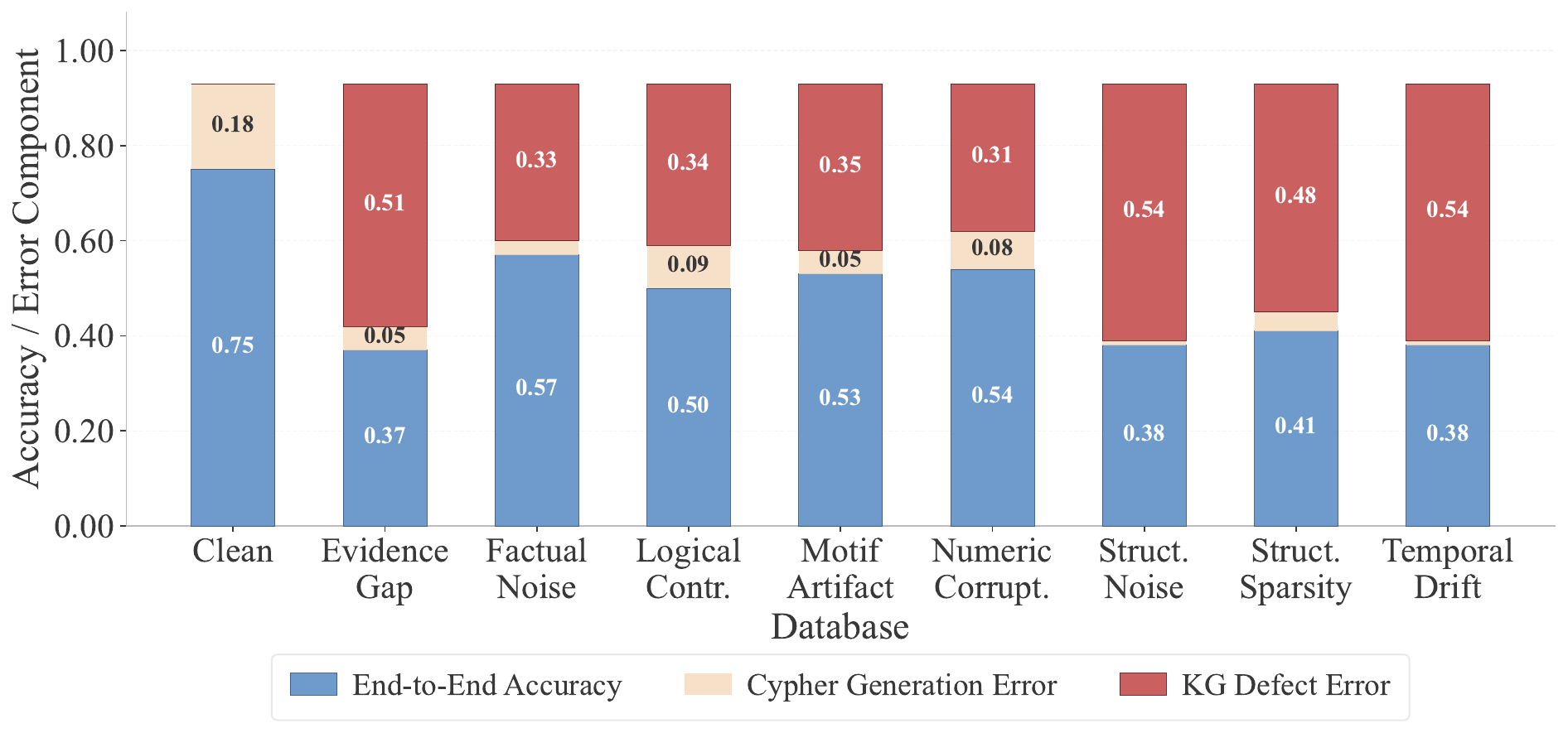}
    \vspace{0.3cm}
    \caption{Error Attribution Stacked Analysis}
    \label{fig3}
\end{figure}

Fig.~\ref{fig3} presents the relative contribution of each error component under different defect scenarios using stacked bars. Based on the decoupled diagnostic framework, the error sources are quantitatively decomposed into reasoning loss, KG defect error, and Cypher generation error. 

The results show that the reasoning loss $\varepsilon_{\text{reason}}$ remains stable at 0.01 across all settings, which is fully consistent with the non-significant result of $p = 0.706$. This demonstrates that, within our experimental context, the intrinsic reasoning capability of the LLM is not the dominant bottleneck of system performance. By contrast, the data quality loss ($\varepsilon_{\text{KG}}$) varies substantially across scenarios. In Structural Noise and Temporal Drift, $\varepsilon_{\text{KG}}$ reaches 0.54, whereas in numeric defect scenarios it is only 0.31, yielding a difference of 74\%. This pattern confirms that the impact of KG defects is strongly dependent on the structural integrity of the database.

The Wilcoxon signed-rank test further verifies the statistical significance of these effects. For all eight defect types, $\varepsilon_{\text{KG}}$ is extremely significant ($p < 0.001$). Among them, the four structural defect types yield large effect sizes: Structural Noise ($d = 1.08$), Temporal Drift ($d = 1.04$), Evidence Gap ($d = 0.98$), and Structural Sparsity ($d = 0.92$). The four numeric defect types show medium effect sizes, ranging from 0.62 to 0.66. In sharp contrast, the reasoning-loss component is completely non-significant ($p = 0.706$, $d = -0.038$), further supporting the conclusion that algorithmic reasoning loss is not the dominant factor.

The Parametric Knowledge Masking Effect is most evident in Evidence Gap ($\varepsilon_{\text{cypher}} = 0.05$) and Structural Sparsity ($\varepsilon_{\text{cypher}} = 0.04$). Relative to $\varepsilon_{\text{cypher}} = 0.18$ on the clean graph, this corresponds to a reduction of more than 70\%. From a Data Integrity and cloud storage management perspective, this phenomenon represents a significant masking effect that obscures underlying data defects. Our results suggest that when database retrieval paths are broken and queries return empty results, the LLM falls back to internal parametric knowledge to generate answers based on semantic priors. It is critical to emphasize that while this occasional correct response partially shrinks the apparent $\varepsilon_{\text{cypher}}$, the overall End-to-End accuracy still declines substantially. More importantly, this mechanism masks the true structural defects in the underlying graph database. 

Consequently, severe data errors may become less detectable for DBAs during routine data quality monitoring. This masking effect implies that conventional monitoring scripts relying solely on end-to-end accuracy are insufficient; decoupled diagnostics are essential to reveal underlying storage corruption. Without a decoupled diagnostic framework, severe storage corruption can remain hidden behind the model's compensatory behavior, compromising the trustworthiness of data integration systems.

\subsection{System Robustness Evaluation}
To evaluate how consistently each query performs under different KG quality conditions, the stability score for each question $q$ is defined as:
\begin{equation}
\text{Stability}(q) = 1 - \sqrt{\frac{1}{N-1} \sum_{i=1}^{N} (s_{q,i} - \bar{s}_q)^2}
\label{eq:stability}
\end{equation}
where $s_{q,i} \in \{0, 0.5, 1\}$ is the judge-assigned score across $N = 9$ databases (one clean and eight defective), and $\bar{s}_q$ is the mean score. Because binary scoring restricts $s_{q,i}$ to at most three values, $\hat{\sigma}_q$ takes only five possible values: 0 (Stability = 1.000), 0.314 (0.686), 0.416 (0.584), 0.471 (0.529), and 0.497 (0.503).

\begin{figure*}[htbp]
    \centering
    \begin{minipage}{0.48\textwidth}
        \centering
        \includegraphics[width=\linewidth]{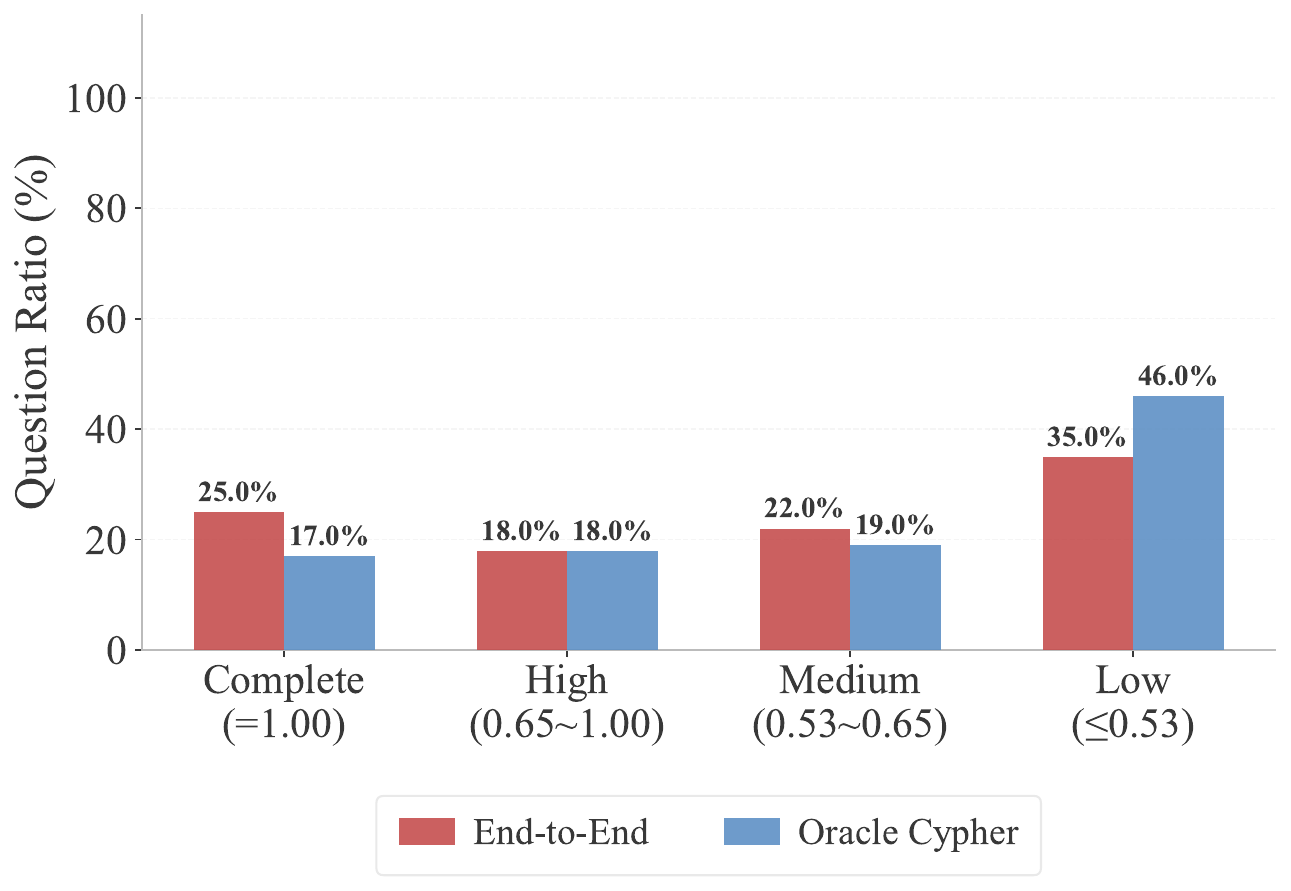}
        \vspace{0.1cm}
        \centerline{(a) Stability Tier Distribution}
    \end{minipage}\hfill
    \begin{minipage}{0.48\textwidth}
        \centering
        \includegraphics[width=\linewidth]{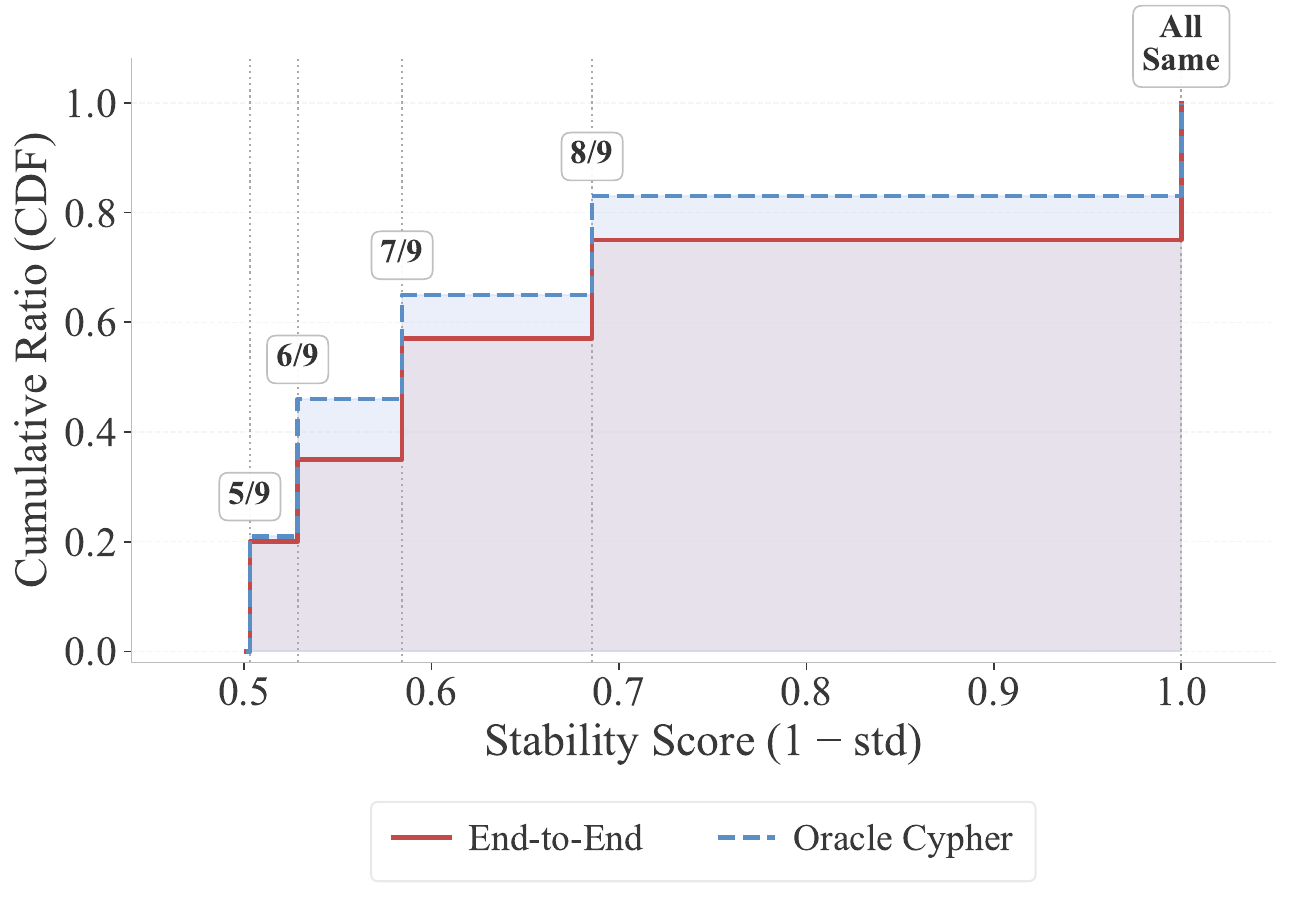}
        \vspace{0.1cm}
        \centerline{(b) CDF of Stability Scores}
    \end{minipage}
    \caption{System Robustness Evaluation (E2E vs. Oracle Cypher)}
    \label{fig4}
\end{figure*}

Fig.~\ref{fig4}(a) compares the question distribution across four stability tiers---Complete (1.000), High (0.686), Medium (0.584), and Low ($\leq 0.529$)---for E2E and Oracle Cypher. Two contrasts are most salient. First, E2E has a higher Complete Stability proportion (25\%) than Oracle Cypher (17\%), indicating that the E2E pipeline contains more queries that superficially remain stable. Second, Oracle Cypher has a higher Low Stability proportion (46\%) than E2E (35\%), reflecting the true, stronger sensitivity of structured database retrieval to KG quality perturbations.

This distribution shift is further visualized by the cumulative distribution function (CDF) in Fig.~\ref{fig4}(b). In the low-score region ($\leq 0.53$), the E2E curve consistently lies below the Oracle Cypher curve, confirming that fewer E2E questions accumulate in the Low Stability range. This pattern strongly aligns with the masking effect discussed previously: the parametric buffering from the LLM partially offsets the instability of broken retrieval paths, shifting outcomes from Low Stability toward Complete Stability and thereby obscuring actual data deterioration.

The Medium Stability and Low Stability questions together account for 57\% of the dataset. Their performance varies substantially across defective graphs, making them the primary targets for future data quality audits and storage robustness optimization in cloud-based graph systems.

\section{Conclusion and Future Work}
This study proposes a decoupled evaluation and diagnostic framework designed to provide quantitative data quality control and integrity monitoring for graph databases during cloud deployment and large-scale multi-source data fusion. By systematically analyzing error sources under eight types of data quality perturbations, three main conclusions are drawn.

First, data asset quality, rather than algorithmic reasoning, emerges as the dominant performance bottleneck within the evaluated Graph-RAG pipeline. For all eight defect types, the data quality loss ($\varepsilon_{\text{KG}}$) reaches an extremely significant level ($p < 0.001$), with a maximum effect size (Cohen's $d$) of 1.08. At a 20\% defect injection rate, system accuracy drops from the baseline of 0.93 to the 0.39--0.62 range, underscoring the critical need for rigorous data quality control.

Second, structural defects (accuracy dropping to 0.39--0.45) cause roughly twice the damage of numeric defects (0.58--0.62). This demonstrates that prioritizing the topological reachability of database storage is significantly more critical than merely correcting numeric node attributes during data ingestion and integration.

Third, the Parametric Knowledge Masking Effect (PKME) poses a critical observability challenge. While algorithmic reasoning loss remains insignificant ($p = 0.706$), severe structural defects cause $\varepsilon_{\text{cypher}}$ to shrink drastically, yielding a Masking Index ($MI$) of over 70\%. Crucially for Database Administrators (DBAs), this masking effect increases the risk of false negatives in automated monitoring scripts, posing a severe challenge for ensuring strict Service Level Agreements (SLAs) in cloud databases. Without a decoupled diagnostic framework, such effects can lead to false confidence in system robustness and undetected data degradation in production environments.

While this study focuses primarily on Cypher-based retrieval, future work will extend in three directions. First, cross-model verification: the audit framework will be deployed across heterogeneous models to evaluate the universality of PKME. Second, cloud system integration: we plan to evolve this static methodology into a Dynamic Quality Assurance System, integrating it directly with cloud-native graph platforms (e.g., Neo4j Aura or Amazon Neptune) for real-time defect interception. Third, spatial scale expansion: the framework will be applied to substantially larger information fusion tasks across the entire Qinghai-Tibet Plateau.

\end{document}